\documentstyle[12pt,epsf]{article}

\begin{document}

\title{\Large\bf Software for Geodynamical Researches \\ Used in the LSGER IAA}
\author{\it Zinovy Malkin, Alexander Voinov, Elena Skurikhina  \\[2mm]
     \normalsize Institute of Applied Astronomy
     nab. Kutuzova 10, St.Petersburg 191187, Russia}
\date{\vspace{-5mm}}
\maketitle

\abstract{
Laboratory of Space Geodesy and Earth Rotation (LSGER) of the
Institute of Applied Astronomy (IAA) of the Russian Academy of
Sciences has been carrying on, since its creation, the computation
of geodynamical products: Earth Orientation Parameters (EOP) and station
coordinates (TRF) based on observations of space geodesy techniques:
Very Long Baseline Interferometry (VLBI), Satellite Laser Ranging (SLR),
Global Positioning System (GPS). Principal software components, used for these investigations, include:
package GROSS for processing of SLR observations, package Bernese for processing of GPS observations,
package OCCAM for processing of VLBI observations, software for data exchange, and
software for combination of space geodesy products.

Package GROSS has been developed in the IAA for
computation of EOP including operational solution with high
accuracy and operativity about 2 days. As a result of special efforts
operational processing of SLR observations is fully automatic.

Package Bernese has been developed in the AIUB and is used
for processing of GPS observations.  Special front-end software
has been developed on the basis of the scripting language Python to
provide easy configuration and automation of processing and
interaction with other packages and libraries.  Computation
of EOP and TRF from global GPS network is planned in the future.

Package OCCAM, originally developed by several West-European and
Russian scientists and modified in the IAA, is mostly used for
processing of VLBI observations for
computation EOP and TRF including operational solution.

To provide automated and reliable data exchange between the IAA
Analysis Center, observation data bases, and International Services
special software was developed.  This software allows to fully automate
downloading of observational data through unstable Internet connections,
data exchange between machines where various components of software used
for processing are installed, etc.

Software for combination of space geodesy EOP and TRF products
is under development.  It is intended to derive combined EOP
and TRF solutions using original VLBI, SLR, and GPS results
to provide independent solution for more valuable contribution
to IERS and other International Services.
}

\footnotetext{Presented at ADASS IX Conference, Hawaii, October 3--6, 1999}

\clearpage
\section{Introduction}

According to IAA mandate in Russian Academy of Sciences,
the Institute is, in particular, responsible for applications
and development of space geodesy methods for geodynamical
researches.
In the framework of this tasks Laboratory of Space Geodesy and Earth
Rotation (LSGER) has been working in the following fields:

\begin{itemize}
\item[---] Development of algorithms and software for
          processing of VLBI, SLR and GPS observations.
\item[---] Investigation and improvement of the dynamic models
          for precise calculations of satellite orbits.
\item[---] Computation of EOP from SLR observations.
\item[---] Computation of EOP, station and radio source coordinates
          from VLBI observations.
\item[---] Computation of EOP and station coordinates from GPS observations.
\item[---] Combining VLBI, SLR, GPS observations for determination of
          EOP and station coordinates.
\item[---] Investigations of regional crust deformation in the framework
          of regional geodynamical projects.
\item[---] Support and investigation of local geodetic networks
          at QUASAR network sites.
\end{itemize}

To support research activity in these topics the LSGER has been using
various software products. Part of them is being designed and advanced
in the Laboratory.

Main software products used in the LSGER are briefly reviewed in this paper.

\section{GROSS package}
\label{gross}

The program package GROSS ({\underline{G}}eodynamics,
{\underline{R}}otation of the Earth, {\underline{O}}rbit
determination {\underline{S}}atellite {\underline{S}}oftware)
is used for analysis of the SLR observations.
The last version of GROSS meets IERS Conventions (1996) with
only exception related to computation of tidal variations
of geopotential.
The package was thoroughly tested and showed high accuracy of
reduction and secondary analysis.
The package is operated on PC under MS DOS/Windows.

\begin{figure}
\epsfxsize=120mm
\centerline{\epsfbox{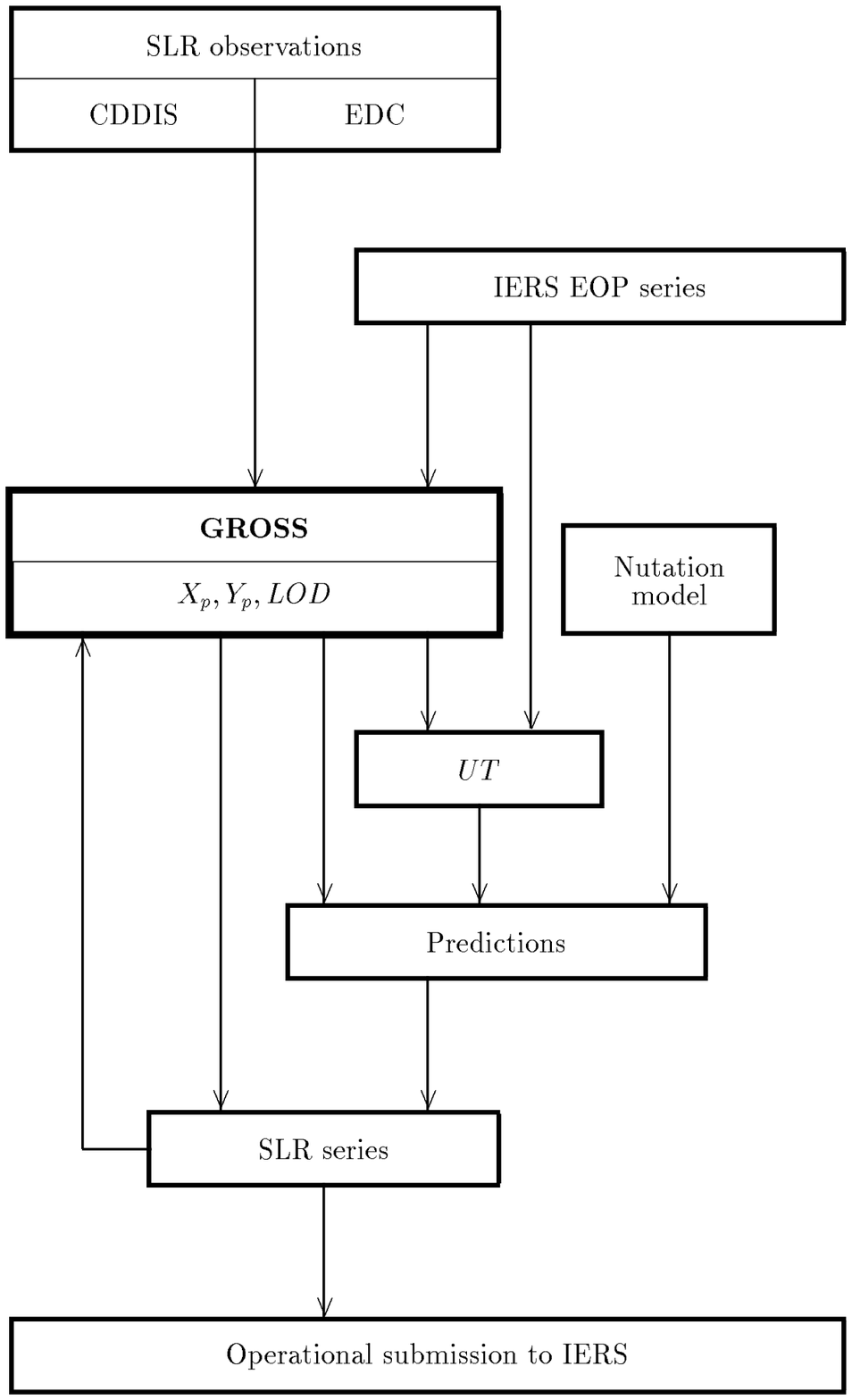}}
\caption{Organization of operational computation of EOP from SLR observations.}
\label{eop_serv}
\end{figure}

Beginning from 1995 the GROSS package is used for analysis of the
observations collected by the global SLR network in the framework
of cooperation with IERS, including participation in the IERS Rapid Service,
and for own investigations.
The special strategy of operational calculations of EOP has been developed
to minimize a gap between last observation and epoch of operational EOP
(now this gap is about 2 days).
Organization of operational computation of EOP from SLR observations is
shown in Fig.~\ref{eop_serv}.
Original method of EOP prediction \cite{iaa_pred} is
used as substantial part of this strategy to provide a priori values
of EOP for latest epochs.

Operational calculations of EOP are being made every working day
fully automatically.
Software used for operational computations is, in fact, distributed
one and includes both MS DOS/Windows (GROSS, data formatting,
supplement service programs, archiving of results) and Unix
(data exchange with world data bases and centers of analysis,
ftp functions) components.
Interaction between Windows and Unix components is realized
via Windows's network SMB protocol implemented on the Unix part by
a {\tt samba} server.

The software works as follows.  Observational data and other
relevant files form Data Centers are automatically downloaded
with desirable time interval (from twice per day to once per week)
on the Unix machine.
These data are picked up by GROSS operating
on Windows machine as everyday scheduled task.
If all needed data are not downloaded in time, GROSS is waiting
for the completion of the data transfer.
Upon the completion of the computation resulting
file is transferred to the Unix machine to be automatically sent to users.
When a problem during data exchange occurred, message is
automatically sent to appropriate persons.
In parallel, EOP files of common IAA use are updated along with
corresponding data base on Windows and Unix machines.
These data are available via anonymous ftp, too.

Before and during computation GROSS controls input data to
prevent wrong results if these data are incomplete or incorrect.
Besides some configuration parameters needed for GROSS
are automatically adjusted to amount and quality of input data.

\begin{figure}[ht]
\epsfxsize=120mm
\centerline{\epsfbox{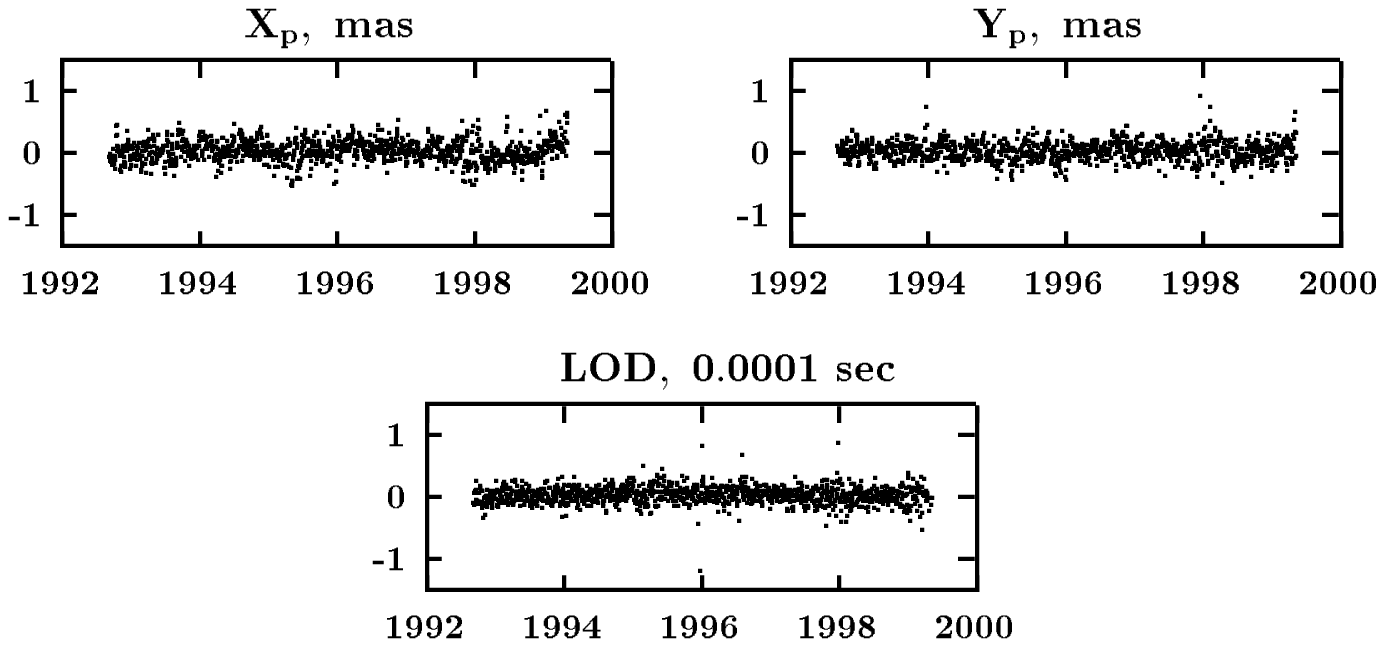}}
\caption{Determination of EOP from the SLR observations.}
\label{fig:slr_dif}
\end{figure}

Both operational and final solutions are
regularly submitted to IERS to be included in IERS operational
and yearly combined solutions.
Fig.~\ref{fig:slr_dif} shows the differences between IAA and IERS solutions.

\section{MAL library}
\label{mal}

Together with the GROSS package the MAL
({\underline{M}}athematical and {\underline{A}}stronomy
{\underline{L}}ibrary) library is being advanced.
MAL contains more than 700 mathematical, astronomical, geodetic,
and service Fortran routines of common use.
Routines are written in Fortran 77 with some Fortran 90 extensions.

Of course, many routines, especially in astronomical and mathematical
parts of MAL provide the same functions as similar routines from other
known libraries, but many of them differ of known ones in efficiency
and/or convenience. Others was included in MAL for completeness.

In opposite to latters, service part of MAL includes mainly original
routines to make programming more comfort.

One of the most serious problem during work on a program package
designed for processing of observations (and not only, indeed)
is its transition between platforms.  It is impossible to
create such a software using only standard set of operators
(at least in Fortran).
Many needed functions such as file operations, reading command line,
getting system information (e.g. date, environment variables), etc,
can be realized only with use of RunTime libraries accompanied
each compiler.  This requires to correct many calls when changing
platform where a package should operate.

To make solution of this problems more simple and without need
of intervention in code of main programs and routines the following
strategy has been using.  All functions required system depending
operators are realized by set of about 30 special system depending
routines with special
names containing \verb"S_" prefix.  For instance, \verb"S_COPY" routine
copies files, \verb"S_GETCL" returns command line, \verb"S_DATE"
returns system date, etc.  In one's turn, these routines contain
calls to RunTime library functions.
Most of "\verb"S_" routines depend only on platform, others
on compiler, too.

Programmer uses these routines in his
software instead of use of RunTime library.  Collection of such
routines for various compilers provides simple installation on
each platform.  Compiling main part of MAL user merely should
add several system depending routines for given platform/compiler
without need to review all package.

It should be mentioned that this strategy does not provide full
transportability.  Some problems still exists, such as use of
direct access file that may have various formats for various
compilers.  Evidently, programmer should keep off use of these
constructions when possible.

Unfortunately, the serious limitation of use of MAL library
is that all documentation is available only in Russian (although
in-code comments are in English).

\section{OCCAM package}
\label{occam}

The OCCAM package \cite{OCCAM34} has been developed
for geodetic VLBI analysis.  It was originated by a group
of West-European scientists (N.~Zarraoa,  H.~Schuh, J. Campbell, P. Stumpff
and others).  Now the package is intensively modified and improved in the IAA
and the St.Petersburg State University.
OCCAM is a compact and mobile package providing high precision
determination of EOP and station coordinates.

The package runs on PC under MS DOS/Windows.
Model of reduction mainly follows IERS Conventions (1996).
The main feature of OCCAM is the use of the Kalman filter technique for
tropospheric zenith delay and clock offset estimation
which are modeled as random walk stochastic process.

\begin{figure}[ht]
\epsfxsize=120mm
\centerline{\epsfbox{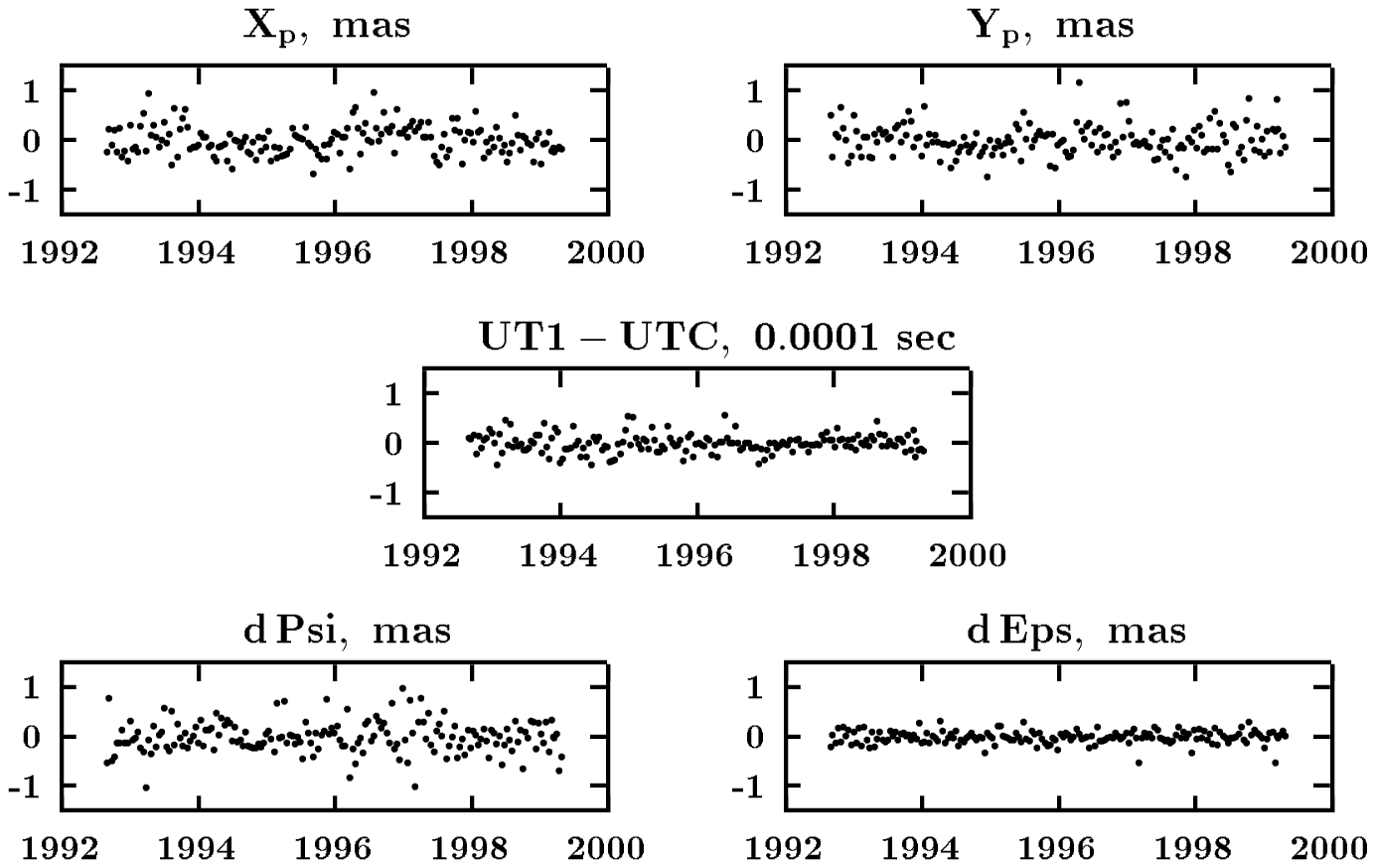}}
\caption{Determination of EOP from the VLBI observations.}
\label{fig:vlbi_dif}
\end{figure}

Beginning from 1997 the package used for routine processing of
VLBI observations collected from the global VLBI network.
At the moment we are processing observations of IRIS, NEOS,
and CORE programs beginning from 1983.
Solutions are regularly submitted to IERS and used for computation
of the IERS combined products.
Cooperation with recently organized International VLBI Service (IVS)
is also very fruitful.
Fig.~\ref{fig:vlbi_dif} shows the differences between IAA and IERS solutions.

\section{Bernese package}
\label{bernese}

Package Bernese has been developed in the Astronomical Institute
of the University of Bern \cite{Bernese}.
It is widely known in the world as one of the most powerful,
flexible, and accurate software for processing of the GPS observations.
Version 4.0 we use at the moment provides computation of EOP,
station coordinates and atmospheric parameters (both troposphere
and ionosphere). Unfortunately, this version is not Y2K-compliant
and should be replaced by a new version before 2000.

We use Bernese mainly for the analysis of observations on regional GPS
networks, such as the Baltic Sea Level project. Routine processing
of the EUREF network is planned.

In process of exploration of Bernese in the LSGER we faced some
problems related to the automation of computation and aggregating
Bernese with other software used for geodynamical researches.

Bernese software consists of the following three subsystems which
reflect three possible levels of user interaction with it:

\begin{itemize}
\item[---] MAIN system: Comprises command line utilities, driven by somewhat
complex ``configuration files'', which either contain or refer to all
necessary information for the processing.

\item[---] MENU system: Interactive modules, helping to specify visually
concrete jobs for the data processing.

\item[---] BPE: Data processing automation system (Bernese Processing Engine).
\end{itemize}

The main ``entry-point'' for a user of Bernese software is MENU system.
Although all the formats of configuration files are documented, their manual
maintenance is cumbersome. Moreover, MENU system performs many essentially
``intelligent'' manipulations with the data, which reflect the professional
experience of the authors of Bernese software.

In the scientific applications of the package, one meets, however,
situations, where the flexibility of both MENU and BPE subsystems is not
sufficient. In particular, many of scientific users of the package have
their own libraries suited for various specific purposes, not covered by
Bernese software.

Direct integration of such libraries with Bernese software seems very
cumbersome and unreliable, especially due to possible interference of
runtime systems.

To solve this problem special software PyGPS was developed.

\section{PyGPS package}
\label{pygps}

PyGPS is a software package aimed at the high level control and automation
of GPS processing with the use of Bernese Software v. 4.0 \cite{Bernese}.
Its core language is Python, some components are written in C and Fortran
and organized as Python extension modules.

In brief, the package consists of:

\begin{enumerate}
\item  Python modules, useful for many purposes, not necessarily related to
Bernese software:

\begin{itemize}
\item[---]  {\tt DateTime}: Julian and GPS date and time operations.

\item[---]  {\tt Rinex}: Reading, writing and converting of RINEX files.

\item[---]  {\tt Marker}: Various operations with GPS sites and their coordinates.

\item[---]  {\tt MAL}: Python interface to the data processing library with
the same name, written in Fortran.
\end{itemize}

\item[---]  Interface to the MAIN subsystem of Bernese software:

\begin{itemize}
\item[---]  General ``wrapper'' for a MAIN program of Bernese software. It
provides means to generate all three kinds of configuration files, needed to
run a Bernese program and to interpret its output.

\item[---]  Specific modules (``classes'') for concrete data processing (or
transformation) utilities: {\tt RXOBV3} (RINEX$\to$Bernese file format conversion),
{\tt CODSPP}, {\tt SNGDIF}, {\tt MAUPRP}, {\tt GPSEST}.

\item[---]  Special modules for selection of ``optimal baseline set'' (see below).
\end{itemize}
\end{enumerate}

\subsection{PyGPS: A Brief Description}

\subsubsection{General Purpose Python Modules}

\paragraph{DateTime:}

This module defines an object with the same name, analogous to the
object {\tt Date} defined above. Date arithmetic is implemented
together with some GPS-oriented methods. Although there is a general
purpose date/time Python extension, written by
M.-A.Lemburg ({\tt http://starship.python.net/crew/lemburg}),
the described one is left for GPS specifics.

\paragraph{Rinex:}

Defines the object with the same name, which implements reading and writing
of RINEX data files and supplies various methods
simplifying splitting and merging of these files according to various
criteria.

\paragraph{MAL:}

This is a wrapping of an applied subroutine package MAL written in Fortran.

\subsubsection{Wrapping the MAIN set of Bernese software utilities}

All programs of the MAIN set of utilities of Bernese software have a lot of
similar input and output properties together with a strategy of their using.
Common input properties include:

\begin{itemize}
\item[---] {\tt *I.INP} files, containing control parameters which affect
  processing;

\item[---] {\tt *F.INP} files, listing the data files to be processed;

\item[---] {\tt *N.INP} files, describing locations of all input and output
  files, which are touched by that program.
\end{itemize}

Common context of running all of these programs, besides creating mentioned
configuration files, may include also creation of an error file, if any, and
notification of caller about the fact of that creation.

The strategy of ``wrapping'' this functionality in the described package is
based upon the notion of Python class, which instance corresponds in that
case to a particular setting of those parameters for the program, which is
represented by that class.

These classes are organized in a hierarchy, which root, the base class
called {\tt MAINPGM} is responsible for the above-described common
features of MAIN programs. It's most important {\em method} is {\tt run},
which performs creation of the configuration files according to the current
attributes of the given class instance and really runs the corresponding
program. After the completion of the program execution, this method also
looks for an error file and displays it to the standard output.

Common functionality, which is inherited by all of the {\tt MAINPGM}
descendants, concludes also in providing convenient means for handling
default and explicit parameters for the corresponding utility. These
parameters may either be {\em class} {\em attributes}, inherited by all its
instances, or default or explicitly overridden arguments of the {\em
constructor} of the class instance. To make dealing with these parameters
more convenient to the user, the notion of class attributes and ``nested
classes'' is used in some particular way.

Wrappers for the main Bernese utilities are implemented as descendants of
that {\tt MAINPGM} class.

A programmer is encouraged to build his/her own class hierarchies, based on
these fundamental classes, which would comprise his/her own experience with
particular sets of processing parameters, occurred suitable in some
particular situations. That way, described package already predefines two
classes for {\tt GPSEST} running, {\tt GPSEST} and {\tt GPSEST\_Final}, the
latter being a descendant of the former, so that the first of them
comprises default parameters, suitable for baseline-wise ambiguity
resolution whereas the second is intended for final least square parameter
estimation.

The original contribution to the processing strategy, accomplished with
PyGPS, concludes in a special iteration loop, aimed at creation of a set of
baselines together with their preprocessing and ambiguity resolution. This
step is described in \cite{FGI:Voinov}.

\section{Software to Transfer Data through Internet over unstable connections}
\label{aftp}

The subroutine package AFTP comprises the following components:
\begin{itemize}
\item[---] Specialized programs to download data from the astronomical
databases USNO, CDDIS, EDC, containing the results of VLBI and SLR
observations.

\item[---] General purpose utilities to exchange data via the ftp protocol.

\item[---] A C/C++ class library, implementing basic operations, used in the
standalone utilities mentioned above.

\item[---] A Python wrapper for this class library.
\end{itemize}

The basic functionality of the package is based upon a finite automaton
schema, which implements robust self-recovering process of background
data transfer over unstable connections. At all points where a new
portion of data or a control protocol response is expected, the
corresponding timeout is checked, and connection is attempted to
reestablish after a moderate pause. This is expressed by the automaton
moving from one state to another.  The "memory" of the described
automaton is the sequence of files yet to download (to be specific let
us describe only the case of downloading data). This queue is automatically
updated during the transfer, so after any system crash the process may be
correctly resumed.

\begin{figure}[ht]
\epsfxsize=100mm
\centerline{\epsfbox{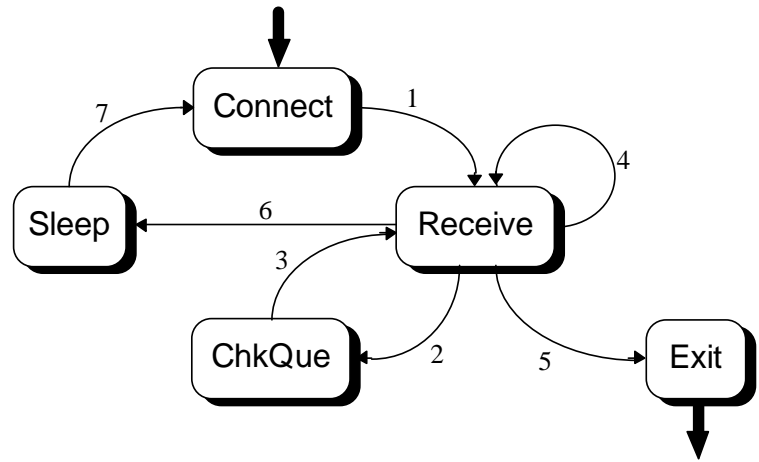}}
\caption{The finite automaton.}
\label{fig:aftp}
\end{figure}

The following states are implemented:
\begin{description}
\item{\tt Connect:}
An attempt to connect to the remote ftp-server. A ``control'' socket is
established, user authentication is done and the remote working directory
is  finally set.  If the response from the server didn't come within a
chosen timeout, the automaton moves to the ``Sleep'' state.

\item{\tt Receive:}
Next data file is downloaded.  The sequence of steps to receive the
contents of this file is also capable to process connection failures.
What is most important in that case, is ability to restart broken
transfer from a point, close to the end of the already downloaded
portion of the file. (About one kilobyte at the end is skipped for safety).

\item{\tt ChkQue:}
The environment, in which the download is performed, is checked for any
events, which may affect the queue of files to download. For example, in
the case of downloading data from the USNO database, its directory is
checked for new files which might appear during the previous piece of
transfer.

\item{\tt Sleep:}
A pause after the connection failure. It's meaning is not to overload the
local CPU by fruitless attempts to reconnect to the remote server.

\item{\tt Exit:}
The queue of file to download is over.
\end{description}

The transition arcs as depicted at the Fig. \ref{fig:aftp}, are as follows:

\begin{enumerate}
\item If a connection attempt was successful, the automaton switches to the
processing of the list of files to download.

\item At the successful completion of the download of the current file, the
automaton either checks the environment as described above, or switches to the
next file to download, if any (the arc (4)).

\item Having completed the checking of asynchronous events the automaton
returns to the processing of the files queue.

\item See (2).

\item The completion of the automaton's functioning.

\item At the connection failure the automaton switches to the "Pause" state.

\item The end of the "Pause".
\end{enumerate}

The practical problems, which are solved with this package, are as follows:
\begin{itemize}
\item[---] Daily monitoring of the USNO database directory.

\item[---] Downloading of the new data, either directly, or with the use of
an intermediate host, an access to which is kindly granted by NASA. This host
is used to repackage the data before downloading. At that respect the
described package represents a multi-agent system (where the agent migration
is not implemented).

\item[---] Downloading of the laser location data from the two databases --
CDDIS and EDC.
\end{itemize}

A universal utility to download an arbitrary collection of files from an
arbitrary ftp server is also implemented. It takes a list of files, where
first two lines describe the server access and the common remote directory
for all the files. This file serves as a "file queue" for the automaton
described above.

\section{Conclusion}

Laboratory of Space Geodesy and Earth Rotation of the Institute
of Applied Astronomy RAS owes full set of software needed to
analysis of modern space geodesy observations -- VLBI, SLR, GPS.
All software is actively used for scientific researches, as a rule
in framework of international programs and projects coordinated
by IERS and IAG providing valuable contribution to join efforts.
High degree of automation of routine computations (especially in
EOP Service) provides everyday and reliable results and saves
many staff's time.

Of course, in spite of limited resources, we are trying to
actively work on developing of software.  The main trends
of this work are:
\begin{itemize}
\item[---] Development of software for combination of original VLBI, SLR,
     and GPS products to derive combined EOP and TRF solutions.
\item[---] Development of OCCAM software and its further
     integration with GROSS package.
\item[---] Development of new independent software for processing of
     GPS and GLONASS (Russian {\underline{GLO}}bal
     {\underline{NA}}vigation {\underline{S}}ysytem) observations
     GRAPE (I.~Gayazov, M.~Ke\-shin).
\item[---] Development of software for integrating various packages
     using chiefly Unix and Python technologies.
\end{itemize}
We hope that this will lead to expanding of field of investigations
and increasing of accuracy of obtained results.


\begin{thebibliography}{9}

\bibitem{iaa_pred}
Malkin Z., Skurikhina E. (1996):
On Prediction of EOP.
Comm. IAA, No 93.

\bibitem{Bernese}
Rothacher M. and Mervart L. (eds) (1996):
Bernese GPS Software Version 4.0.
Astronomical Institute, University of Berne.

\bibitem{OCCAM34}
Titov O., Zarraoa N. (1997):
OCCAM 3.4 User's Guide.
Comm. IAA, No 69.

\bibitem{FGI:Voinov}
Voinov (1999):
GPS data processing automation with the use of Python
scripting language.
In: M.Poutanen, J.Kakkuri (eds.), Final results of the Baltic Sea
Level 1997 GPS Campaign. Rep. Finn. Geod. Inst., 1999, {\bf 99:4}, 75--84.

\end{thebibliography}
\end{document}